\documentclass[pagesize,12pt,a4paper]{scrartcl}
\usepackage{german}
\usepackage{amsmath}
\usepackage{amsthm}
\usepackage{amssymb}
\usepackage{amsxtra}
\usepackage[squaren]{SIunits}
\usepackage{graphicx}
\usepackage{braket}
\usepackage{capt-of}
\usepackage{ae}
\usepackage{wasysym}
\usepackage{makeidx}
\usepackage{inputenc}
\usepackage[T1]{fontenc}
\usepackage{url}
\usepackage{eepic}
\usepackage{mathrsfs}
\usepackage{comment}
\usepackage{mparhack}
\newcommand{\ir}{\mathrm{i}}
\newcommand{\e}{\mathrm{e}}

\newcommand{\eins}{{\mathbf 1}}

\newcommand{\phys}{\text{phys}}
\renewcommand{\jmath}{j}

\newcommand*\idx[1]{\index{#1}#1}

\providecommand{\norm}[1]{\lVert#1\rVert}

\DeclareMathOperator{\dv}{d}

\DeclareMathOperator*{\diag}{diag}
\ifx\KOMAScript\undefined%
  \DeclareRobustCommand{\KOMAScript}{\textsf{K\kern.05em O\kern.05em%
      M\kern.05em A\kern.1em-\kern.1em Script}}
\fi
\newlength{\help}
\newlength{\minuslaenge}
\newtheoremstyle{note}
  {3pt}
  {3pt}
  {\rmshape}
  {}
  {\bfseries}
  {:}
  {.5em}
  {}

\theoremstyle{note}

\makeatletter

 \def\vec#1{\ensuremath{\mathchoice
                     {\mbox{\boldmath$\displaystyle\mathbf{#1}$}}
                     {\mbox{\boldmath$\textstyle\mathbf{#1}$}}
                     {\mbox{\boldmath$\scriptstyle\mathbf{#1}$}}
                     {\mbox{\boldmath$\scriptscriptstyle\mathbf{#1}$}}}}%
\makeatother

\begin{document}

  \title{Heisenberg Algebra and String Theory}
  \author{Norbert Dragon and Florian Oppermann\\
          Institut f\"ur Theoretische Physik\\
          Leibniz Universit\"at Hannover \\
orcid 0000-0002-3809-524X
}
\date{}

\maketitle

\begin{abstract} 
If the algebra of the Poincar generators is enlarged by the spacetime position operator~$X=(X_0,\dots, X_{D-1})$ 
then the spectra of the momentum $P$ and the mass $P^2$ are unbounded and continuous. 
In particular, the constraint $(P^2 - m^2)\Psi_{\phys}=0$ of the covariant string 
has no solution in the space which admits $X$: All physical states vanish, $\Psi_{\phys}=0$.
Vice versa, a space spanned by mass eigenstates does not admit the position operator~$X$ in $D$ dimensions.

A massless particle does not allow a spatial position operator $\vec X$.

The domain of Heisenberg pairs $X^i$ and $P^j$, $i,j\in \set{1,\dots D-2}$, $D > 2$, which commute with $P^+=(P^0 + P_z)/\sqrt{2}$\,,  $[P^+,X^i] = 0$\,,
does not allow for a space with massless or tachyonic states, which is
mapped to itself by rotations, leave alone Lorentz transformations. 
This is true in all dimensions  and makes the algebraic calculation 
of the critical dimension, $D=26$, of the  bosonic string meaningless: the light cone string is not Lorentz invariant.

\end{abstract}

\newpage


\section*{In Medias Res}

\subsection*{Restriction of States}

The bracket notation of quantum states $\ket{p,i}_{\mathcal M}$, $i\in I \subset {\mathbb N}$, 
is to be read in the distributional sense that the states are
\begin{equation}
\label{ketnotation}
\ket{\Psi} = \sum_{i\in I} \int_{\mathcal M}\! \dv\! p \, \Psi^i(p)\, \ket{p,i}_{\mathcal M}
\end{equation}
where each wave function
\begin{equation}
\Psi: \left \{ 
\begin{array}{clc}
\mathcal M \times I &\rightarrow & \mathbb C \\
(p,i) &\mapsto &\Psi^i(p)
\end{array}
\right . 
\end{equation}
is measurable and square integrable with respect to some volume form $\mu$. We denote the states 
of the Hilbert space $\mathcal H(\mathcal M)$ simply by their wave functions $\Psi:\mathcal M\times I \rightarrow \mathbb C$. Their scalar product is
\begin{equation}
\label{scalprod}
\braket{\Phi|\Psi}= \sum_{i\in I} \int_{\mathcal M}\! \dv\!\mu_p \, \Phi^{i\,*}(p)\, \Psi^i(p)
\end{equation}
with a positive volume form $\dv\!\mu_p=\mu(p)\dv\!p$. Usually one omits the label $\mathcal M$ of the distribution $\ket{p,i}_{\mathcal M}$
if the domain of the test functions is self understood. For $\mathcal M \subset \mathbb R^D$, however, we have to distinguish 
the distribution $\ket{p,i}_{\mathcal M}$ from the distribution $\ket{p,i}_{\mathbb R^D}$.

For our purposes the notation of states as equivalence classes of wave functions is superior to the 
bracket notation which 
tempts to mistake a state valued distribution for a state. Smooth (i.e.  infinitely often differentiable), rapidly decreasing wave functions constitute an important
dense subspace, the Schwartz space  $S(\mathcal M)\subset\mathcal H(\mathcal M)$. If a finite dimensional Lie group~$G$ is unitarily represented by
$U_g:\mathcal H(\mathcal M)\rightarrow \mathcal H(\mathcal M)$, $U_g U_{g'}= U_{g\,g'}$, $g,g'\in G$, where $\mathcal M = G/H$
is the $G$-orbit  of an arbitrarily chosen point $\underline p\in \mathcal M$ with stability group $H\subset G$, 
then $S(\mathcal M)$ is the domain $D(\mathcal A)$ of the polynomial algebra $\mathcal A$
of the skew hermitian generators of~$U_g$ and is invariant under~$U_g$ \cite{schmuedgen}.
In bracket notation smoothness and rapid decrease are usually disregarded as they are no properties of the distribution $\ket{p,i}$ but of the test functions.

Eq. (\ref{ketnotation}) shows that states are equivalence classes of functions $\Psi$. They are equivalent if the support 
of their difference vanishes (has measure zero), because by definition each Lebesgue integral vanishes for integrands of vanishing support.

Hence the restriction of a state $\Psi\in \mathcal H(\mathbb R^D)$ to a mass shell $\mathcal M\subset \mathbb R^D$ of vanishing $D$-dimensional measure
\begin{equation}
\label{restriction}
\Psi_{\mathcal M}: p \stackrel{?}{\mapsto}
\left \{ 
\begin{array}{clc}
0 &\text{if} &p \ne \mathcal M\\
\Psi(p) &\text{if} & p \in \mathcal M\\
\end{array}
\right . 
\end{equation}
yields zero if considered as a state in $\mathcal H(\mathbb R^D)$ because the support of the restricted function vanishes.
Considered as state in the different Hilbert space $\mathcal H(\mathcal M)$ (this is no subspace of $\mathcal H(\mathbb R^D)$) the restriction
is undefined as the equivalence class of each $\Psi \in \mathcal H(\mathbb R^D)$ consists of functions with whatever restriction to $\mathcal M$ you choose.
Restriction of wavefunctions to a submanifold $\mathcal M$ of measure zero does not define a projection of states from $\mathcal H(\mathbb R^D)$ 
to $\mathcal H(\mathcal M)$. In this respect a discrete argument~$i$ and a continuous argument $p$ of a wave function differ essentially.

The problem seems avoided in the dense subspace of smooth wave functions (which constitute no Hilbert space as their Cauchy sequences need not
converge to smooth functions). Each has a unique smooth representative. 
Its restriction is unique and defines a wave function in $\mathcal H(\mathcal M)$. But
to restrict a state to a submanifold is no continuous map of $\mathcal H(\mathbb R^D)$. No matter how small (measured with the norm in $\mathcal H(\mathbb R^D)$)
the difference of~$\Psi$ and a smooth state is, $\Psi$'s restriction to $\mathcal M$ is arbitrary. The problem is similar to
a function of the rational numbers which is undefined on irrational numbers. It does $\emph{not}$ define a real function.

In bracket notation the restriction of $\ket{p,i}_{\mathbb R^D}$ to momenta in $\mathcal M$ seems no big deal:
Just pick the subset $p\in \mathcal M$. But there is nothing to select! The symbol $\ket{p,i}_{\mathbb R^D}$ denotes a distribution 
which has a value at test functions $\Psi:\mathbb R^D\times I \rightarrow \mathbb C$, not at momenta. The token $p$
does not stand for an independent variable but is the integration variable in (\ref{ketnotation}): Like a summation variable it has no value but ranges over the domain of integration.

\subsection*{Generators of Unitary Representations}

Each $\omega$  from the Lie algebra~$\mathfrak g$ of a finite dimensional Lie group $G$
generates by the exponential map the elements $g_t=\e^{t\,\omega}$, $t\in \mathbb R$,
of a one parameter group, $g_t g_{t'}=g_{t+t'}$. 

The skew hermitian generators $-\ir M_\omega$ of the one parameter subgroups of a unitary representation $g \mapsto U_g$ , 
$U_g:\mathcal H(\mathcal M)\rightarrow \mathcal H(\mathcal M)$, $U_g U_{g'} = U_{g\, g'}$, 
represent $\mathfrak g$ on a subspace of states $\Psi$ on which all $U_{\e^{t\omega}}$ act differentiably 
\begin{equation}
-\ir M_\omega \Psi = \lim_{t\rightarrow 0} (U_{\e^{t\omega}}\Psi - \Psi) /t\ .
\end{equation}

They generate $U_{\e^{t\,\omega}}=\e^{-\ir\, t M_\omega}$ by their spectral resolution~$E_\lambda$ (which depends on $\omega$)
\begin{equation}
M_\omega = \int\! \dv\! E_{\lambda}\,\lambda\ ,\ 
U_{\e^{t\,\omega}} = \e^{-\ir\,t\, M_\omega} = \int\! \dv\! E_{\lambda}\,  \e^{-\ir\,  t\,\lambda}\,\ .
\end{equation}
This definition of $U_{\e^{t\,\omega}}$ extends to the complete Hilbert space $\mathcal H$  while the power series $\sum_k (-\ir t M_\omega )^k \Psi/k!$ converges only 
for states $\Psi$ from some analytic subspace which is unsuitably small for some purposes, e.g. it contains no
functions of compact support.

If for all $\Psi\in \mathcal H(G)$ the map $g \mapsto U_g \Psi $ from $G$ to $\mathcal H(G)$ is measurable, then the G\aa rding space exists which is 
spanned by states which, using  an invariant volume form~$\dv\! \mu$, are averaged with smooth
functions $f: G \rightarrow \mathbb C$ of \emph{compact} support
\begin{equation}
\label{smoothing}
\Psi_f = \int_{G}\! \dv\! \mu_g \, f(g)\, U_g \Psi\ .
\end{equation}
These integrals over measurable integrands of compact support exist. 

The G\aa rding space is dense in $\mathcal H(G)$ \cite{dixmier}. Its spanning states transform by
\begin{equation}
\label{smoothtrans}
U_g \Psi_f = \Psi_{f\circ g^{-1}}\ .
\end{equation}
The G\aa rding space is mapped to itself by all $U_g$ and their generators $M_\omega$. 
Applied to smooth states the products $U_{g(t_1 \dots t_n)}=U_{\e^{t_1\omega_1}}\cdots U_{\e^{t_n\omega_n}}$ depend  differentiably
on all~$t_i$. So the derivatives $\partial_{t_1}\dots \partial_{t_n}U_{g(t_1\dots t_n)_{|_{t=0}}} = (-\ir)^n M_{\omega_1}\cdots M_{\omega_n}$ exist
no matter how large~$n$ is. Applied to smooth states not only the products $M_\omega M_{\omega'}$ and $ M_{\omega'} M_\omega$ exist 
such that $M_\omega M_{\omega'}-  M_{\omega'} M_\omega =\ir  M_{[\omega,\omega']}$,
but the polynomial algebra $\mathcal A$ of the generators $M_\omega$ \cite{schmuedgen}.
Their composition, the product of the polynomial algebra, is defined because the subspace $D(\mathcal A)$ of smooth states, the domain of $\mathcal A$, 
is invariant under all $U_g$ and $M_\omega$.

That the G\aa rding space exists and constitutes a common, dense and invariant domain of the polynomial algebra of the generators
is highly welcome to physicists who manipulate the unbounded generators algebraically not caring about domains. 
That its states are smooth in the orbit $G/H$ makes differential geometry and topology applicable, because smooth states
in Hilbert space correspond uniquely to smooth functions. 

Only operators which map smooth states to smooth states are acceptable in relativistic theories.
Otherwise they yield unphysical states with divergent expectation values of momentum and angular momentum. 
This unphysical property is behind the mathematical singularities which 
exclude,  as we show below, the light cone string. 

Remarkably, the averaging (\ref{smoothing}) yields states, which are are not only smooth but also rapidly decreasing for large momenta and energies.

\section*{The Heisenberg Algebra}

The momentum $P=(P^0,\dots P^{D-1})$ generates the unitary representation $U_a = \e^{\ir\, P \,a},$ $a\in \mathbb R^D$, of 
translations in spacetime. If one enlarges their algebra 
by commuting, hermitian position operators \mbox{$X=(X_0,\dots X_{D-1})$}
which are translated
\begin{equation} 
\label{xtrans}
\e^{\ir\, a\, P} X_n \,\e^{-\ir\, a\,  P}= X_n - a_n\ 
\end{equation}
then functions $f(X)$ such as $V_b(X ) = \e^{\ir\,b\, X}$, are shifted, $U_a f(X) U_a^{-1}= f(X-a)$, 
$U_a V_b U_a^{-1}=V_b(X-a) = V_b\, \e^{-\ir\, a\, b}$. 
These are the Weyl relations 
\begin{equation}
\label{weyl}
U_a V_b =V_b U_a\, \e^{-\ir\, a\, b}\ ,\ U_a U_b =U_{a+b}\ ,\ V_a V_b =V_{a+b}\ ,\ a,b \in \mathbb R^D\ .
\end{equation}
The products $U_a\, V_b\, \e^{\ir\, c}$
constitute the $(2D+1)$-dimensional Heisenberg group. 

By (\ref{xtrans}), $X$ and $X-a$ are unitarily equivalent for all $a\in \mathbb R^D$: The $X$-spectrum is translation invariant.
The same holds for $P$: Solving the Weyl relations for $V_b U_a V_b{}^{-1}= U_a \,\e^{\ir a\,b}$
and differentiating, $\partial_{a_n}$ at $a=0$, shows
\begin{equation}
\label{ptrans}
\e^{\ir\,b\,X}P^n\e^{-\ir\,b\,X}= P^n+b^n\ . 
\end{equation}
The spectra of $P$ and $P+b$ coincide for all $b$: The  $P$-spectrum is translation invariant.

By the Stone-von Neumann theorem \cite[Theorem XI.84]{reed3} (for $D=1$)
each unitary representation of the Weyl relations is unitarily equivalent to the one in a Hilbert space~$\mathcal H$
of states $\Psi: p \mapsto \Psi(p)$
which map $p\in \mathbb R^D$ almost everywhere to $\Psi(p)$ in some Hilbert space~$\mathcal N$. The unitary representation acts multiplicatively and by translation
\begin{equation}
\label{heisenrep}
\begin{gathered}
(U_a \Psi)(p)= \e^{\ir\, a\, p}\,\Psi(p)\,,\ (V_b \Psi)(p)= \Psi(p+b)\ ,\\ 
\braket{\Phi | \Psi}_{\mathcal H}= \int\! \dv^D\!\!p \braket{\Phi(p)|\Psi(p)}_{\mathcal N}\ .
\end{gathered}
\end{equation}
In case that a second pair $P'$, $X'$ exists which commutes with the first pair and acts in~$\mathcal N$, one applies the theorem 
to $\mathcal N$ and the second pair and extends it to $D=2$  and by further induction to $D$ momenta $P^m$ and~$D$ commuting 
position operators $X_n$, if they exist.

If $\mathcal N=\mathbb C$, then the representation is irreducible.
It is generated by the operators 
\begin{equation}
\label{stoneN}
\eins \Psi = \Psi\ ,\ 
\bigl (P^m\Psi\bigr)\,(p) = p^m \Psi(p)\ ,\ 
\bigl (X_n\Psi\bigr)\,(p) = -\ir \partial_{p^n} \Psi(p)\ , 
\end{equation}
which are hermitian with respect to the translation invariant measure $\dv^D\!p$ (\ref{heisenrep}) and 
satisfy the Heisenberg Lie algebra
\begin{equation}
\label{heisenberg}
[\eins, X_m] = 0 = [\eins, P^m] \ ,\ [P^m, P^n]=0 = [X_m, X_n]\ , \ [ P^n,X_m]= \ir\delta^n{}_m\eins\ .
\end{equation}

Their polynomial algebra $\mathcal A$ is defined on and maps to itself the \idx{Schwartz space} $S(\mathbb R^D, \mathcal N)$ of 
smooth functions $\Psi:\mathbb R^D \rightarrow \mathcal N$ which together with each of their derivatives decrease rapidly.
For each pair of multiindices $\alpha,\beta\in \mathbb N_0{}^D$ 
there exists a bound $C_{\alpha,\beta,\Psi}$ such that for all $p\in \mathbb R^D$
\begin{equation}
\label{schwartz}
\norm{p^\alpha \partial_{\beta} \Psi(p)}_{\mathcal N} < C_{\alpha,\beta,\Psi}\ .
\end{equation}
Here $p^\alpha$ denotes $p_1^{\alpha_1}\, p_2^{\alpha_2}\dots p_D^{\alpha_D} $ and
$\partial_\beta = (\partial_{p^1})^{\beta_1}\, (\partial_{p^2})^{\beta_2}\dots (\partial_{p^D})^{\beta_D} $.

Recall that an eigenvalue of an operator $O$ is a number $\lambda$, for which $(O-\lambda)\Phi=0$ has a solution $\Phi\ne 0$. The spectrum of~$O$ is 
the set of numbers $\lambda$, for which $(O-\lambda)$ has no bounded inverse.
A number $\lambda$ picked from a continuous spectrum need not be an eigenvalue.

A Heisenberg paired momentum $P$, $[P,X]=\ir$, has no eigenstate. 
A state $(P-\lambda) \Phi = 0 $, $\Phi\ne 0$, in the domain of the
polynomial algebra of $X$ and $P$ would imply the contradiction
\begin{equation}
0-0=\braket{(P-\lambda)\Phi|X\Phi}- \braket{\Phi|X(P-\lambda)\Phi}=\braket{\Phi|[P,X]\Phi }=\ir\,\braket{\Phi|\Phi}\ne 0\ .
\end{equation}

Using $D$ Heisenberg pairs $X_m$ and $P^n$ one easily specifies  generators $M_\omega = \omega^{mn} M_{mn}/2$ 
of Lorentz transformations $U_{\e^\omega} = \e^{-\ir M_{\omega}}$ \cite{thooft},
\begin{equation}
\label{lorentzcov}
-\ir M^{mn}\Psi \stackrel{?}{=} \ir (P^m X^n - P^n X^m)\Psi + \Gamma^{mn} \Psi\ , 
\end{equation}
where $\Gamma^{mn}$ are skew hermitian matrices which commute with $X$ and $P$ and represent the Lorentz algebra 
\footnote{We use units of $\hslash = 1$ and the metric in an orthonormal basis $\eta = \diag (1,-1,\dots, -1)$.}
\begin{equation}
\label{loralgebra}
 [\Gamma^{mn}, \Gamma^{rs}] = -\eta^{mr}\Gamma^{ns}+\eta^{ms}\Gamma^{nr}+\eta^{nr}\Gamma^{ms}-\eta^{ns}\Gamma^{mr}
\end{equation}
as do the differential operators $l^{mn}= \ir (P^m X^n - P^n X^m)$.
Nonvanishing $\Gamma^{mn}$ can occur in case the scalar product is indefinite.
\emph{However:} If $X^0$ exists, then for each $\Psi$ with an energy probability density 
supported in a bounded, positive range, there are suitable $b$ and states $\e^{\ir\,b\,X^0}\Psi$ in which one is certain to measure
negative energies only.


\section*{Models}

In covariant string theory and also in simpler models with the pretentious title \lq relativistic particle\rq\  \cite{witten, hanson, scherk}
the space of one-particle states $\mathfrak L^2(\mathbb R^D) \otimes \mathcal N$ 
factorizes into the space $\mathfrak L^2(\mathbb R^D)$ of square integrable functions of $\mathbb R^D$, acted upon by $D$ Heisenberg pairs $X_m$ and~$P^n$,
and~$\mathcal N$,
a denumerable sum of finite-dimensional eigen\-spa\-ces of a level operator $N$,
$[N,X_m]=[N,P^n]=0$, with a nondegenerate, not necessarily definite scalar product. 

By the Weyl relations (\ref{ptrans}), $P$ has the purely continuous, unbounded spectrum~$\mathbb R^{D}$ and 
$P^2= P^mP^n\eta_{nm}$ has the purely continuous, unbounded spectrum $\mathbb R$
while one-particle states in phenomenologically acceptable theories have nonnegative energy 
and are (sums of) mass eigenstates with  discrete masses. 
This discrete mass distinguishes one-particle states from many-particle states whose relative motion imparts their invariant mass $P^2$
a continuous, nonnegative spectrum. 

Therefore and by reasons of the classical model one requires the physical states $\Psi_\phys$ of a relativistic quantum particle
with $D$ Heisenberg pairs to satisfy the constraint
\begin{equation}
\label{constraint}
\bigl (P^m P^n \eta_{mn} - m^2(N)\bigr)\Psi_{\text{phys}}=0
\end{equation}
possibly with different masses for the different components of $\Psi_\phys$.

The constraint is \emph{utterly wrong.} 

The constraint has no solution as the support of 
\begin{equation}
\Psi_{\text{phys}}: p\stackrel{?}{\mapsto}
\left \{ 
\begin{array}{clc}
0 &\text{if} &p^2 \ne m^2\\
\Psi_{\text{phys}}(p) &\text{if} & p^2 = m^2\\
\end{array}
\right . 
\end{equation}
is a set of va\-ni\-shing $D$-dimensional measure.
Thus its scalar product (\ref{heisenrep})
vanishes for all $\Phi \in \mathcal H(\mathbb R^D)$. Each physical state in $\mathcal H(\mathbb R^D)$ vanishes,
\begin{equation}
\label{physnul}
\Psi_\text{phys}= 0\ . 
\end{equation}
The covariant string and the pretended 'relativistic particles' have no physical states which a constraint could select. 

This flaw had been observed previously \cite{bahns, dimock, grundling} but without the explicit conclusion
that such a contradiction is serious. We disagree: Whether a space of nonvanishing, physical states exists is fundamental.

The other way round: If the $(2D+1)$-dimensional Heisenberg group existed in a space spanned by mass eigenstates (as postulated in \cite{dimock}) and 
if this group commuted with~$m^2$ (as it does in string theory)
then the  domain $D(\mathcal A)$ of the algebra $\mathcal A$ of its generators $X^n$ and $P^m$ decomposed into spaces of definite
mass and contained an eigenstate~$\Psi$ of mass~$m$. By standard argu\-ment~$\Psi$ is orthogonal to all states $\Phi$ of different mass $m'\ne m$, 
\begin{equation}
-(m'^2- m^2)\braket{\Phi|\Psi}=\braket{(P^2 - m'^2)\Phi | \Psi}- \braket{\Phi | (P^2 - m^2)\Psi}= 0\ .
\end{equation}
For all states $\Phi$ of the \emph{same} mass~$m$ the hermiticity of $P^n$ and the constraint imply 
\begin{equation}
\begin{aligned}
\label{comxp}
0 - 0 &= \braket{(P^2 - m^2) \Phi | P^n X_n \Psi}- \braket{\Phi | P^n X_n (P^2 - m^2)\Psi } \\
&= \braket{\Phi | [ P^2 - m^2, P^n X_n]\Psi }= 2\ir \braket{\Phi | P^2 \Psi } = 2\ir\, m^2 \braket{\Phi | \Psi }\ .
\end{aligned}
\end{equation}
So massive states $\Psi\in D(\mathcal A)$ are orthogonal to all states, $\braket{\Phi | \Psi}=0$, and vanish
\begin{equation}
\label{nuldisc}
\Psi=0\ .
\end{equation}
For massless states we repeat the argument (\ref{comxp}) with $P^n X_n$ replaced by $P^0 X_0$ and obtain
\begin{equation}
\braket{P^0 \Phi | P^0 \Psi}=0
\end{equation}
for all massless $\Phi$ in $D(\mathcal A)$.
This again implies $\Psi=0$ because in $D(\mathcal A)$ the scalar product $\braket{P^0\cdot  |P^0\cdot  }$ exists and is
nondegenerate. 

Each state in the domain of the algebra  $\mathcal A$ vanishes: The polynomial algebra of $D$ pairs $P^m$ and $X_n$ does not exist
in a space spanned by mass eigenstates.

The argument did not use the self-adjointness of $X$ nor  the positivity $\braket{\Psi |\Psi}\ge 0$ of the scalar product but only 
its nondegeneracy: If $\braket{\Phi | \Psi}=0$ for all $\Phi$ then $\Psi = 0$.

The absence of a $D$-component  position operator $X$ Heisenberg paired with the momentum $P$ does not prohibit local fields $\phi(x)$. The field's argument $x$
is not the spectral value of a position operator: For each open domain $\mathcal U \subset \mathbb R^D$, no matter how bounded, the field operators 
\begin{equation}
\label{phif}
\phi_f = \int_{\mathcal U}\! \dv^D\!x\, f(x)\, \phi(x)\ ,
\end{equation}
applied to the vacuum, create according to the Reeh-Schlieder theorem \cite{reeh} a dense set of one-particle states
as $f$ ranges over the smooth functions with support contained in a fixed $\mathcal U$. So a field $\phi(x)$ does not create or annihilate particles at the
position~$x$. 

That one anyhow talks about fields $\phi(x)$ creating or annihilating particles at $x$ is casual language about virtual particles at vertices
of Feynman diagrams as if they and their position were observable.

Neither massive nor massless particles allow a Heisenberg pair $[P^0, X^0]=\ir$ as $X^0$  would generate shifts
$\e^{\ir\,b\,X^0}\,P^0\,\e^{-\ir\,b\,X^0}=P^0 + b$. But the spectrum of $P^0=\sqrt{m^2+\vec P^2}$ is non-negative and not translation invariant.
Otherwise, as we had already argued, for each state~$\Psi$ with an energy probability density supported in a bounded, positive 
domain there would exist states $\e^{\ir\,b\,X^0}\Psi$ with energy probability density in a range of negative energies.

The same argument prohibits for massive and massless particles the pair $[P^+, X^-]=\ir$ with shifts $\e^{\ir\,b\,X^-}$  because the 
lightlike momentum $P^+=(P^0 + P_z)/\sqrt{2}$ is non-negative and not translation invariant.
Hence, if operators~$X^0$ or $X^-$ exist at all, then for a tachyon at most.  

That $X^0$ and $X^-$ do not exist should not be embellished as \lq $X^0$ is not a good operator\rq\  or 
\lq $X^0$ is illegal\rq or \lq $X^0$ is not gauge invariant\rq\ or whatever the names are to 
slander an operator while not giving up on it completely. One has to state clearly: $X^0$ and $X^-$ are disallowed, they do not exist.
Having accepted their non-existence one has to check all properties which one had derived previously while assuming
their existence.

The pedigree of a theory does not help against logic: if a set of mathematical relations hold true, then its consequences hold true,
no matter whether the relations were obtained from classically consistent constraints, from  canonical quantization of a mere gauge condition or whether they occur
in a two-dimensional conformal field theory. Relations do not hold \lq in the sense of conformal field theory\rq, they hold or they hold not.
In a relativisticc theory we consider them a contradiction, if they exclude a unitary representation the Poincaré group in Hilbert space.

\subsection*{Massive Particle}

Because $\mathcal H(\mathcal M)$ does not allow  $D$ Heisenberg pairs the Lorentz generators are not given by 
(\ref{lorentzcov}) as assumed for the relativistic string or the \lq relativistic particle\rq .
But they are well known. 
In the massive case, $m > 0$, they act on momentum wave functions on which matrices 
$\Gamma_{ij}= - \Gamma_{ji}$,  $i,j \in \set{1,\dots D-1}$, generate a representation of SO$(D-1)$, \footnote{
We use matrix notation 
and suppress indices of the components of $\Psi$ and~$\Gamma_{ij}$} 
\begin{equation}
\mathcal M_m  = \set{p:p^0 = \sqrt{m^2 + \vec p^2}}\subset \mathbb R^D\ ,\ 
(P^n \Psi)(p) = p^n \Psi(p)\ ,
\end{equation}
\begin{equation}
\begin{aligned}
\label{massiv}
\bigl(-\ir M_{ij}\Psi\bigr)(p) & = -\bigl(p^i \partial_{p^j} - p^j \partial_{p^i}\bigr)\Psi(p) + \Gamma_{ij}\Psi(p)\ ,\\
\bigl(-\ir M_{0i}\Psi\bigr)(p) & = p^0\,\partial_{p^i}\Psi(p) + \Gamma_{ij}\frac{p^j}{p^0 + m}\Psi(p)\ .
\end{aligned}
\end{equation}
Verifying the Lorentz algebra observe $\sum_{i=1}^{D-1} p^ip^i=(p^0)^2 - m^2=(p^0+m)(p^0-m)$. The gene\-ra\-tors are skew-hermitian with respect to
the invariant measure $\dv\!\mu_p = (\dv^{D-1}\!p)/p^0$.

As massive, positive energy particles do not allow a timelike Heisenberg pair $P^0, X_0$ or a lightlike pair $P^+, X_+$, they allow at most $D-1$
spacelike Heisenberg pairs. Their position operators span a spacelike plane with a unit normal vector, which can be considered 
to denote the $D$-velocity of a set of associated observers at relative rest who measure position with the operator $\vec X$. 
On wave functions $\hat\Psi = \Psi/(m^2+\vec p^2)^{1/4}$ 
with translation invariant measure $\dv^{D-1}\!p$ the position operators act by $-\ir \partial_{p^i}$. On  the wave functions $\Psi$
with the Lorentz invariant measure $\dv^{D-1}\!p/\sqrt{m^2+\vec p^2}$ they act as
\begin{equation}
\ir\, X_i \Psi(p) = \partial_{p^i}\Psi(p) - \frac{p^i}{2\,(m^2+\vec p^2)}\Psi(p)\ .
\end{equation}
One easily confirms that the operators $X_i$ are defined in and leave invariant the space of smooth, rapidly decreasing wave functions $\Psi \in S(\mathbb R^{D-1})$, 
the domain of the polynomial algebra~$\mathcal A$ of the Poincaré generators $\vec P$, $P^0=\sqrt{m^2 + \vec P^2}$ and 
$M_{mn}$ (\ref{massiv}). 

Massive particles \emph{do allow} a position operator $\vec X$ with $(D-1)$ commuting components, which transforms under rotations as a vector
and constitutes $(D-1)$ Heisenberg pairs with the spatial momentum $\vec P$.

\subsection*{Massless Particle}

Not so well known is the representation of the Lorentz generators 
on massless states \cite{bose, dragon, fronsdal,lomont},
\begin{equation} 
\label{masselosnord}
\begin{aligned}
\bigl(-\ir M_{ij}\Psi\bigr)_N(p)&= - \bigl(p^i\partial_{p^j} - p^j\partial_{p^i}\bigr)\Psi_N(p) + h_{ij}\,\Psi_N(p)\ ,\\
\bigl(-\ir M_{zi}\Psi\bigr)_N(p)&= - \bigl(p_z\partial_{p^i} - p^i\partial_{p_z}\bigr)\Psi_N(p) + h_{ik}\frac{p^k}{|\vec p|+p_z}\,\Psi_N(p)\ ,\\
\bigl(-\ir M_{0i}\Psi\bigr)_N(p)&=  |\vec p|\partial_{p^i}\Psi_N(p) + h_{ik}\frac{p^k}{|\vec p|+p_z}\,\Psi_N(p)\ ,\\
\bigl(-\ir M_{0z}\Psi\bigr)_N(p)&=  |\vec p|\partial_{p_z}\Psi_N(p)\ , 
\end{aligned}
\end{equation}
where $i,j,k\in \set{1,\dots D-2}$, $p_z = p^{D-1}$ and $h_{ij}= - h_{ji}$, $h_{ij}^\star = - h_{ij}$, generate a representation of SO$(D-2)$.
In checking note $p^k p^k = |\vec p|^2- p_z^{\,2}=(|\vec p|+ p_z)(|\vec p|- p_z)$. 
The generators are skew-hermitian with respect to the invariant measure $\dv\!\mu_p = \dv^{D-1}\!p/|\vec p|$.

The subscript $N$ signifies that the smooth function $\Psi_N$ represents the smooth state
$\Psi$ only in the northern, open coordinate patch $\mathcal U_N$ outside the negative $p_z$-axis,
\begin{equation}
\mathcal U_N=\set{p: p^2 = 0\,,\,|{\vec p}|+ p_z > 0}\ ,\ \mathcal U_S=\set{p: p^2 = 0\,,\,|{\vec p}|- p_z > 0}\ .
\end{equation}
Within $\mathcal U_N$ the function $1/(|\vec p|+ p_z)$ and the generators $M_{mn}$ are smooth.

In the intersection $\mathcal U_N\cap \mathcal U_S$ (outside the $p_z$-axis) $\Psi_N$ is related to
$\Psi_S= h_{SN}\Psi_N$ by a transition function $h_{SN}$ \footnote{A 
detailed discussion of the bundle structure is submitted.}, for example
in $D=4$ with $h_{ij}=-\ir h \varepsilon^{ij}$, $2h \in \mathbb Z$,
\begin{equation}
\label{suednordwell}
\Psi_S(p) = h_{SN}(p)\,\Psi_N(p)\ ,\ h_{SN}(p) = \e^{2\ir\, h\varphi(p)}=\Bigl (\frac{p_x+\ir p_y}{\sqrt{p_x^2+p_y^2}}\Bigr)^{2h}\ .
\end{equation}
Both $\Psi_S$ and $(M_{mn}\Psi)_S= h_{SN} (M_{mn} \Psi)_N$ are smooth in $\mathcal U_S$, 
\begin{equation}
\label{masselossued}
\begin{aligned}
\bigl(-\ir M_{12}\Psi\bigr)_S(p)&= - \bigl(p_x\partial_{p_y} - p_y\partial_{p_x}\bigr)\Psi_S(p) + \ir\, h\,\Psi_S(p)\ ,\\
\bigl(-\ir M_{31}\Psi\bigr)_S(p)&= - \bigl(p_z\partial_{p_x} - p_x\partial_{p_z}\bigr)\Psi_S(p) - \ir\, h\,\frac{p_y}{|\vec p| - p_z}\Psi_S(p)\ ,\\
\bigl(-\ir M_{32}\Psi\bigr)_S(p)&= - \bigl(p_z\partial_{p_y} - p_y\partial_{p_z}\bigr)\Psi_S(p) + \ir\, h\,\frac{p_x}{|\vec p| - p_z}\Psi_S(p)\ ,\\
\bigl(-\ir M_{01}\Psi\bigr)_S(p)&=  |\vec p| \partial_{p_x} \Psi_S(p) +\ir\,h\,\frac{p_y}{|\vec p| - p_z}\Psi_S(p)\ ,\\
\bigl(-\ir M_{02}\Psi\bigr)_S(p)&=  |\vec p| \partial_{p_y} \Psi_S(p) -\ir\,h\,\frac{p_x}{|\vec p| - p_z}\Psi_S(p)\ ,\\
\bigl(-\ir M_{03}\Psi\bigr)_S(p)&=  |\vec p| \partial_{p_z} \Psi_S(p)\ .
\end{aligned}
\end{equation}
Massless particles \emph{do not allow} a position operator $\vec X$: 
If one enlarges the algebra of the Poincaré generators by Heisenberg partners $X_j$ of the spatial momenta, $[P^i, X_j] = \ir\, \delta^i{}_j$, 
$i,j\in\set{1,\dots D-1}$, 
then this algebra contains
\begin{equation}
P^0=\sqrt{\vec P^2}\ ,\  
[X_j,[X_j, P^0]]=-\frac{D-2}{|\vec P|}
\end{equation}
and, for $D > 2$, all powers of  $1/|\vec P|$. To be in the domain of the algebra, the wave functions have 
to decrease near $\vec p = 0$ faster than any power of $|\vec p|$. As the domain is invariant under $V_{\vec b}=\e^{\ir \, \vec b\,\vec X}$ also all
$(V_{\vec b}\Psi)(\vec p)=\Psi(\vec p+ \vec b)$ have to vanish at $\vec p=0$ for all~$\vec b$, thus $\Psi(\vec b)=0$ everywhere: There is no common, invariant domain 
of $\sqrt{\vec P^2}$, $(D-1)$ Heisenberg pairs $P^i, X_j$ and their Heisenberg group. A massless particle does not allow for 
the abelian translation group  of its spatial momentum and its generators $\vec X$.


Contrary to the rule of thumb, a point in a spectrum can be essential.
Not only integrals are important in quantum physics but also the domain of smoothness, orbits and fixpoints.  
Different from massive particles the spectrum of the spatial momentum of massless particles contains a Lorentz fixpoint, 
the momentum  $\vec p = 0$ where $p^0=\sqrt{\vec p^2}$ is not smooth. This distinguished point breaks the translation invariance of the
spectrum of $\vec P$. So there cannot exist position operators which generate translations of spatial momentum. 
The fixpoint of Lorentz transformations in the massless shell explains the failure of all attempts \cite{wightman} to construct a position operator 
for massless particles.

This outcome disappoints expectations, because we see the world and reconstruct the position of all objects by light
which we receive as massless quanta. 
But we see massive objects, not the photon. Rather we see by means of photons 
which we annihilate in our retina.

\subsection*{Tachyon}

The light cone string employs massless and massive states
and tachyon states $\Psi$ with momenta $p\in \mathcal M_{\text{Ta}}$\,,
\begin{equation}
p = (E, \sqrt{\mu^2 +E^2}\,\vec n)\ ,\ E\in \mathbb R\ ,\ \vec n=(n^1,\dots, n^{D-2}, n_z) \in S^{D-2}\ .
\end{equation}
It postulates \cite[page 23]{arutyunov} transverse Heisenberg pairs $P^i$, $X^j$, $i,j \in \set{1,\dots D-2}$ which commute with
$P^+ = (P^0 + P_z)/\sqrt{2}$ 
\begin{equation}
\label{heisentrans}
 [P^i, X^j]= -\ir \delta^{ij}\ ,\ [P^i, P^j]= 0\ ,\  [X^i, X^j] = 0 \ ,\ [X^i, P^+] = 0
\end{equation}
and imply for tachyonic 
\begin{equation}
P^- = (P^0 - P_z)/\sqrt{2} = (-\mu^2 + \sum_{i=1}^{D-2} P^iP^i)/(2 P^+)\ ,
\end{equation}
no matter whether the operators $X^i$ are self-adjoint or not,
\begin{equation}
\label{einsdurchpplus}
2 P^- + \ir\, \sum_{j=1}^{D-2} [ P^j X^j, P^-] = -\frac{\mu^2}{P^+}\ .
\end{equation}
The momentum $P^+$ vanishes on the cylinder $\mathcal C\subset{\mathcal M_{\text{Ta}}}$ with radius $\mu$, 
\begin{equation}
\mathcal C = \set{q: q^2 = - \mu^2\ \wedge\   q^+ = 0}= \set{q: q = (E, \mu\,\vec n', -E)\ ,\ E\in \mathbb R\ ,\  \vec n' \in S^{D-3}}\ .
\end{equation}

Consequently the operators $X^j$ can be applied only to states $\Psi$ which vanish on $\mathcal C $
\footnote{in the sense that in no $\mathcal M_{\text{Ta}}$-neighbourhoud of a $q\in \mathcal C$ the modulus $|\Psi(p)|^2$ is larger almost everywhere than some positive $c$.}.
Otherwise, if $|\Psi(p)|^2 > c > 0$ almost everywhere in a neighbourhood of some $ q\in \mathcal C$, then
$(1/P^+)\Psi$ is not square integrable: In each (sufficiently small) neighbourhood of $q$ there are coordinates $(y^1,\dots y^{D-2},p^+)$ of the tachyon shell such that
the Lorentz invariant measure $\dv\!\mu_p > c' \dv^{D-2}\!y\dv\!p^+$ with $c'>0,$ and $|1/P^+ \Psi)(y,p^+)|^2> c/|p^+|^2$.

But the condition that $\Psi$ vanish on $\mathcal C$ excludes the unitary action of the group of rotations $R$
\begin{equation}
(U_R \Psi)(R p) = \Psi(p)
\end{equation}
because rotations act transitively on $S^{D-2}$. For each momentum $p$ there are rotations $R$ with 
\begin{equation}
(R \vec n) = (\mu \vec n', - E)/\sqrt{\mu^2+ E^2}\ ,\ \vec n'\in S^{D-3}\ , 
\end{equation}
which rotate $p=(p^0,p^1,\dots,p_z)$ to $R p \in \mathcal C$. So the rotated states $U_R\Psi$ are all in the domain of $1/P^+$ only if $\Psi=0$.

Slightly varied the argument applies also to massless states for which we use 
\begin{equation}
\sum_{j=1}^{D-2}[X^j,[X^j, P^-]] = -\frac{D-2}{P^+}\ .
\end{equation}
There $\mathcal C$ consists of momenta $(E,0\dots,-E)$
with spatial momentum on the negative $z$-axis. In cylindrical coordinates with 
$r = \sqrt{\sum_{i=1}^{D-2}p^ip^i}$ and
$\dv^{D-1}\!p = r^{D-3}\dv\! r \dv\! p_z \dv\! \Omega$ the function
$(1/P^+)^{(D-2)/2} \Psi$ is not square integrable if  $|\Psi(p)|^2 > c > 0$ almost everywhere
in a neighbourhood $\mathcal U$ of some $q\in \mathcal C$. (By the triangle inequality 
$\sqrt{2}p^+=|\vec p|-|p_z|< r$ for $p_z < 0$.) But each spatial momentum 
can be rotated into the negative $z$-direction. So also the massless particles of the light cone string do not allow for a domain
of the polynomial algebra of $P$ and transverse $X^i$ (which commute with $P^+$) (\ref{heisentrans}), which is invariant under rotations.

The algebraic calculation that the canonically quantized Lorentz generators of the light cone string satisfy the Lorentz algebra only 
in the critical dimension $D=26$ is meaningless: 
in no dimension $D > 2$ is there a dense domain of the polynomial algebra of the operators $X^i,\,P^j$ and $P^-$ which is invariant under rotations,
leave alone the Lorentz group.

The failure of supposed generators, motivated by canonical quantization or otherwise, to satisfy the Lorentz Lie algebra unless $D=26$
only disproves the assumed generators. Different generators (\ref{massiv}) of the Poincaré group
exist in all dimensions if and only if the spin multiplets exist, on which the matrix generators of SO$(D-1)$ act.
A proof of $D=26$ would have to show that for each $D>3$ and $D\ne 26$ there is a mass level at which the SO$(D-2)$ multiplets 
of the light cone string fail to constitute SO$(D-1)$ multiplets while in $D=26$ complete SO$(25)$ multiplets exist on all massive levels. 


\section*{Conclusion}
\lq It is always possible that unknown to myself I am up to my neck in quicksand and sinking fast.\rq\,\cite{fischler}

The covariant quantum string has no physical one-particle states (\ref{physnul},\ref{nuldisc}). 

Canonical quantization of the light cone string yields an algebra which contains $1/P^+$ (\ref{einsdurchpplus}).
It excludes a unitary representation of the Lorentz group.
The formal calculation that in $D=26$ canonically quantized generators satisfy the Lorentz algebra is meaningless as in no dimension
do they generate a unitary representation of the group. 

To draw conclusions is the task of the  reader which he cannot pass to his likes or dislikes. 

It is a presently unsolved task to describe what string theory is about. 
Its operators do not allow a dense domain of one-particle states with a unitary representation of the Poincaré group.
Whether string theory can possibly be reinterpreted as dealing with virtual particles, which do not have to 
exist in a Hilbert space, remains to be seen.

\section*{Acknowledgements}
Norbert Dragon thanks Gleb Arutyunov for an e-mail correspondence about (\ref{comxp}) and Wilfried Buchmüller, Stefan Theisen and Sergei Kuzenko for extended, 
clarifying discussions.

\end{document}